\documentclass[aps,prl,preprint,superscriptaddress,showpacs,floatfix]{revtex4-1}
\usepackage{setspace}
\usepackage{amsmath,amssymb}
\usepackage{graphicx}
\usepackage{dcolumn}
\usepackage{bm}
\usepackage{multirow}

\begin{document}
\title{Universal Scalings in 2D Anisotropic Dipolar Excitonic Systems}
\author{Chern Chuang}
\author{Jianshu Cao}
\email{jianshu@mit.edu}
\affiliation{Department of Chemistry, Massachusetts Institute of Technology, MA 02139, USA}
\date{\today}

\begin{abstract}
Low-dimensional excitonic materials have inspired much interest owing to their novel physical and technological prospects. In particular, those with strong in-plane anisotropy are among the most intriguing but short of general analyses. We establish the universal functional form of the anisotropic dispersion in the small $k$ limit for 2D dipolar excitonic systems. While the energy is linearly dispersed in the direction parallel to the dipole in-plane, the perpendicular direction is dispersionless up to linear order, which can be explained by the quantum interference effect of the interaction among the constituents of 1D subsystems. The anisotropic dispersion results in a $E^{\sim0.5}$ scaling of the system density of states and predicts unique spectroscopic signatures including: (1) disorder-induced absorption linewidth, $W(\sigma)\sim\sigma^{2.8}$, with $\sigma$ the disorder strength, (2) temperature dependent absorption linewidth, $W(T)\sim T^{s+1.5}$, with $s$ the exponent of the environment spectral density, and (3) the out-of-plane angular $\theta$ dependence of the peak splittings in absorption spectra, $\Delta E(\theta)\propto\sin^2\theta$. These predictions are confirmed quantitatively with numerical simulations of molecular thin films and tubules.
\end{abstract}

\maketitle

\textit{Introduction.}---The discovery of graphene heralded the emergence of a plethora of novel two-dimensional (2D) materials with wide ranging exotic properties that attract interests of both fundamental physics and technological prospects.\cite{Neto2009RMP,Novoselov2012Nature} Apart from the inorganic 2D materials,\cite{David2014ACSNano,RevModPhys.90.021001,Sergei2018JPCL,Mohite2018NatComm,LuReview2019Small,Libai2020NatComm} organic excitonic systems possess distinct advantages with high tunability and low processing cost.\cite{Jasper2005PRL,Paltiel2013SciRep,Bardeen2014,Kuhn2015PhysRep,CheeKong,SpanoACR2017,Ginsberg2019NatMat,Brixner2020ChemSci} Generically speaking, these systems are particularly interesting owing to the incommensurability essential to 2D excitonic systems with anisotropic dipolar interactions is nonexistent in their 1D counterparts where the signs of the couplings are by construct homogeneous.\cite{JasperPRL2015,Chem2019} These characteristics can be found not only in dipolar excitonic materials, which is the main focus of this Letter, but also other systems such as optical lattices of two-level atoms\cite{Gross2017Science}, trapped ions,\cite{Blatt2008} or Rydberg gas.\cite{RevModPhys.82.2313}



As the combination of the reduced dimensionality, anisotropy, and long-range interactions leads to burgeoning active research, it also poses a challenge.  In fact, most of the existing theoretical analyses are restricted to limiting cases such as nearest-neighbor (NN) or isotropic $r^{-3}$ couplings.\cite{SchreiberToyozawa1982,Malyshev2D,JeremyNJP,Arend2D} On the other hand, while computational studies are capable of revealing physical properties within specific material systems,\cite{Heine2014,Thygesen2018} obtaining fundamental insights universally applicable remains difficult. These results understandably have limited applicability, and a general treatment of anisotropic 2D systems is lacking. 

In this Letter we aim at filling in this gap by providing a scaling analysis of anisotropic 2D dipolar excitonic systems. Here, we focus on the long-range contribution to the dipole coupling and derive scaling relations in the low quasimomentum regime of the exciton dispersion. This scaling regime reflects coarse-grained measures of the transition dipole, consequently the resulting spectroscopic signatures are universal regardless of atomistic details. We further investigate the exciton density of states (DOS) at low energy and examine its spectroscopic consequences that can be readily tested experimentally. To corroborate the universality of our predictions, we also compare with realistic models of molecular films and tubules and find quantitative agreement.\\

\begin{figure}[t]
	\centering
  \includegraphics[width=16cm]{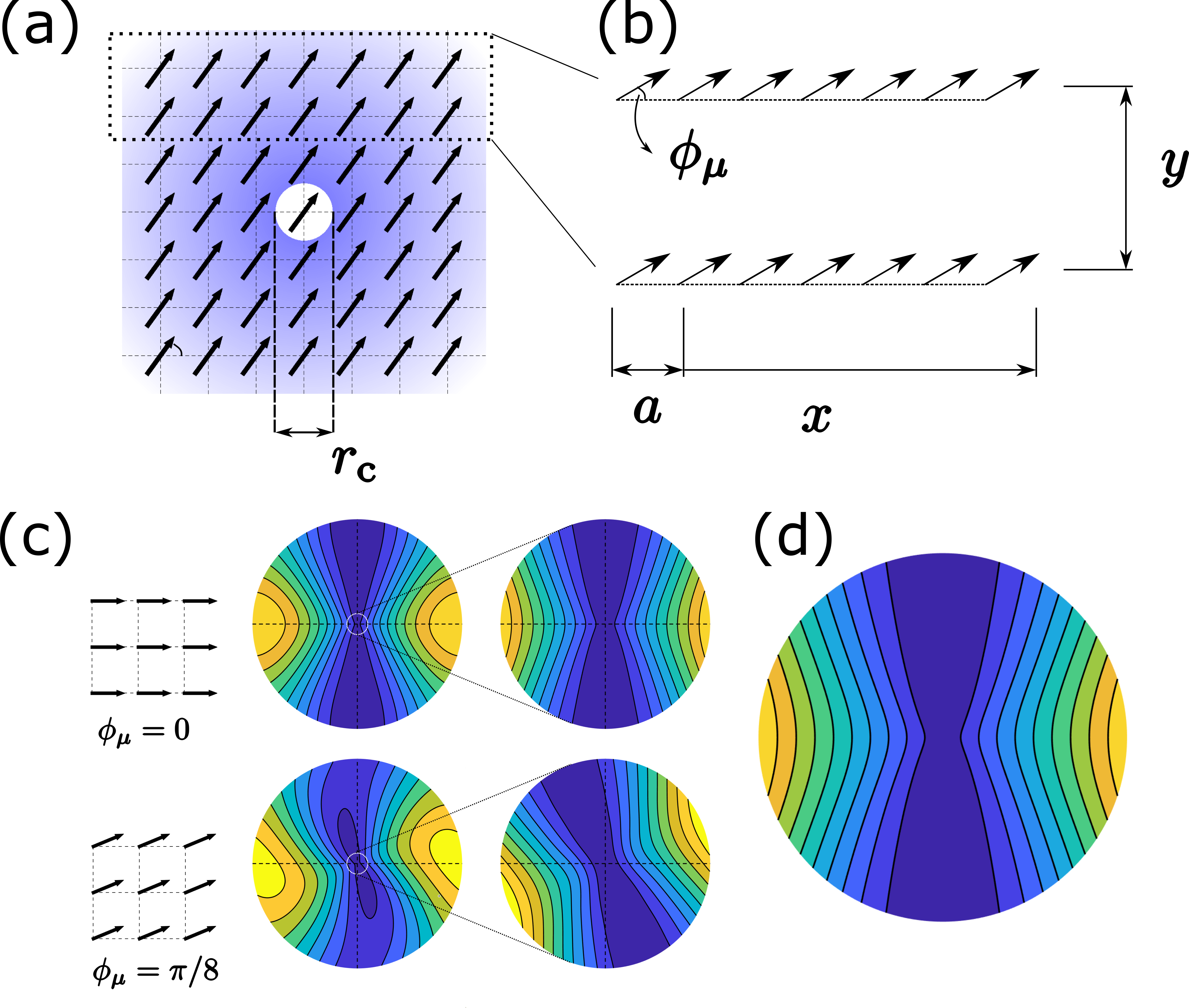}
  \caption{(a) A 2D square lattice of in-plane dipoles. (b) Two parallel chains of dipoles. (c) Left to right: Square lattices with different dipole orientations ($\phi_\mu=0,~\pi/8$), their 2D dispersion contours ($k\le\frac{\pi}{a}$), and those zoomed in ($k\le\frac{\pi}{10a}$). The dashed lines indicate the (reciprocal) lattice vectors. 
  (d) Dispersion of the continuum model predicted by Eq.~(\ref{eqn:continuum_k2}) with $k\le\frac{\pi}{10r_\mathrm{c}}$ and $r_\mathrm{c}=a$.}
  \label{fig:dispersion}
\end{figure}

\textit{Continuum description of 2D excitonic systems.}---We consider a translationally invariant 2D lattice, as shown in Fig.~\ref{fig:dispersion}(a), where at each site $\vec{r}=r(\cos\phi_r,\sin\phi_r,0)$ we assign a dipole $\vec{\mu}_r=\mu_0(\cos\theta_\mu\cos\phi_\mu,\cos\theta_\mu\sin\phi_\mu,\sin\theta_\mu)$. The interaction between the origin and an arbitrary site is given by $J(\vec{r})=\frac{\mu_0^2}{r^3}[1-3\sin^2\theta_\mu\cos(\phi_r-\phi_\mu)]$. Applying the Bloch theorem, the energy dispersion relation reads
\begin{eqnarray}
E(\vec{k})&=&E_0+\sum_{\vec{n}\neq0}J(\vec{r}_n)e^{i\vec{k}\cdot\vec{r}_n},\label{eqn:FourierSum}
\end{eqnarray}
where $\vec{k}$ is the quasimomentum of the corresponding wavefunction. For isotropic NN-coupled square lattices, this gives rise to the conventional cosine dispersion and a constant DOS. 

Taking the continuum limit of Eq.~(\ref{eqn:FourierSum}), which becomes exact in the limit of $r\rightarrow\infty$, and assuming a circular cutoff at $r=r_\mathrm{c}$, represented as the shaded region in Fig.~\ref{fig:dispersion}(a), we obtain
\begin{eqnarray}
E_\mathrm{c}(\vec{k})&=&\frac{1}{a_0}\int d\vec{r}e^{i\vec{k}\cdot\vec{r}}J(\vec{r})\nonumber\\
&=&\frac{\mu_0^2}{a_0}\int_{r_\mathrm{c}}^\infty \frac{dr}{r^2}\int_0^{2\pi}d\phi_r\cdot e^{ikr\cos(\phi_k-\phi_r)}[1-3\sin^2\theta_\mu\cos^2(\phi_r-\phi_\mu)]\nonumber\\
&=&\frac{\mu_0^2}{a_0}\left[\left(1-3\sin^2\theta_\mu\sin^2\Delta_k\right)I_0(k)-3\sin^2\theta_\mu\left(1-2\sin^2\Delta_k\right)I_2(k)\right]\label{eqn:continuum_exact}
\end{eqnarray}
where $a_0$ is the unit cell area, $\vec{k}=k(\cos\phi_k,\sin\phi_k,0)$, and $\Delta_k=\phi_\mu-\phi_k$ is the angle between vectors $\vec{k}$ and $\vec{\mu}$. Here $r_\mathrm{c}$ serves as an adjustable parameter on the order of $\pi r_\mathrm{c}^2\approx a_0$. Its specific value depends on the lattice configuration, dipole orientation, and the form of excitonic coupling, $J(\vec{r})$, especially in the short-range regime. In the last equation we integrate over $\phi_r$ and obtain the closed form expressions of $I_0(k)$ and $I_2(k)$ in terms of Bessel functions, as given in S1 of Supplemental Material (SM). Eq.~(\ref{eqn:continuum_exact}) can be used to numerically calculate the continuum model exactly, see S3 of SM. 

We note that it is precisely the long-range nature of the dipolar coupling that makes the continuum treatment feasible, and thus one should focus on the small $k$ regime and the scaling properties therein.
A series expansion of Eq.~(\ref{eqn:continuum_exact}) up to the order of $k^2$ yields
\begin{eqnarray}
E_\mathrm{c}(\vec{k})&=&\bar{E}\left\{\left(2-3\sin^2\theta_\mu\right)+\right.\nonumber\\
&&\left.\left[-2+2\sin^2\theta_\mu(2-\sin^2\Delta_k)\right]\cdot|kr_\mathrm{c}|+\right.\nonumber\\
&&\left.\left[\frac{1}{2}+\frac{3\sin^2\theta_\mu}{8}\left(-3+2\sin^2\Delta_k\right)\right]\cdot|kr_\mathrm{c}|^2\right\}+\mathcal{O}(k^4)
\label{eqn:continuum_k2}
\end{eqnarray}
where $\bar{E}=\frac{\pi\mu_0^2}{a_0r_\mathrm{c}}$. It can be shown that all odd-order terms vanish except for the linear one, such that Eq.~(\ref{eqn:continuum_k2}) is accurate up to $k^3$ compared to Eq.~(\ref{eqn:continuum_exact}). Eq.~(\ref{eqn:continuum_k2}) is the main result of the current contribution.\cite{fn1} We shall focus on discussing the implications of Eq.~(\ref{eqn:continuum_k2}) for the in-plane dipole configuration ($\theta_\mu=\pi/2$) in the following.

Surprisingly, the leading term in Eq.~(\ref{eqn:continuum_k2}) is linear. This can be understood in terms of coherent summation of the dipole vectors resulting in constructive interference in the direction parallel to the dipole and destructive interference in the perpendicular direction, elaborated in the next section. We note that similar linear dispersion has been recently reported in monolayer MoS$_2$ and is attributed to a different type of physics associated with the inter- and intra-valley exchange interactions.\cite{MacDonald2015PRB,Louie2015PRL,Lan2021ACSNano}
The quadratic term is similar to the standard dispersion of NN-coupled models and serves as a background that is more isotropic in comparison.\cite{fn2} 

We note that the continuum model does not account for the discrete contribution in Eq.~(\ref{eqn:FourierSum}), which is dominantly short-ranged. As such, the continuum model is universal and does not depend on the discrete lattice parameters such as the primitive vectors of the underlying Bravais lattice or other atomistic details. This is especially true for the linear term since the leading correction from the short-range contribution omitted here is $k^2$.\cite{fn3} We confirm this by comparing the numerically exact dispersion of two representative square lattices with different $\phi_\mu$'s, shown in Fig.\ref{fig:dispersion}(c), with that of Eq.~(\ref{eqn:continuum_k2}), Fig.\ref{fig:dispersion}(d). The full dispersion [Fig.\ref{fig:dispersion}(c), middle column] reflects the change of the dipole lattice geometry but their low-$k$ counterparts [Fig.\ref{fig:dispersion}(c), right column] agree well with Eq.~(\ref{eqn:continuum_k2}).

\textit{Quantum Interference of Constituent 1D Strips.}---
A simple physical understanding can be achieved by adopting an intuitive approach to interpreting the Fourier sum Eq.~(\ref{eqn:FourierSum}). Here the dispersion $E(\vec{k})$ is obtained by first decomposing a 2D lattice into an array of 1D strips along the direction perpendicular to $\vec{k}$, then Fourier transforming the coherent couplings between the strips along $\vec{k}$.
\begin{enumerate}
\item Consider first the special case where $\vec{k}$ is parallel to $\vec{\mu}$, \textit{i.e.} $\Delta_k=0$. Essentially this is equivalent to calculating the coherent coupling between 1D strips of perpendicular dipoles, a configuration illustrated by taking $\phi_\mu=\pi/2$ in Fig.~\ref{fig:dispersion}(b). In the continuum limit, the magnitude of the coherent sum of such couplings is linearly proportional to the length of the strips, and has a $r^{-2}$ dependence on the inter-strip separation $r$.\cite{Chuang2014} As a result, the leading contribution in the Fourier basis scales linear in $k$: $E(k) = 2\int dr \cos(k r)\cdot r^{-2}=2 k\cdot\int d(kr) \cos(kr)\cdot(kr)^{-2}\propto k$.
\item The other extreme case is $\Delta_k=\pi/2$, where $\vec{k}\cdot\vec{\mu}=0$ and illustrated by taking $\phi_\mu=0$ in Fig.~\ref{fig:dispersion}(b). In stark contrast, successive cancellation of aligned dipoles leads to vanishing dipole strength of the 1D strips and, thus, vanishing inter-strip couplings.\cite{Barford2007,Chuang2014} Consequently, the dispersion is flat to linear order in $k$.
\item Generally speaking, for an arbitrary angle $\Delta_k$ there is constructive interference between the perpendicular components and destructive interference between the parallel ones. This leads to the $|\vec{k}|\cdot\cos^2\Delta_k$ term in Eq.~(\ref{eqn:continuum_k2}), with derivation detailed in S2B of SM. 
\end{enumerate}
We emphasize that such effects disappear in the incoherent, classical limit, such as those studied in the literature of F\"{o}rster energy transfer, where dipole summation is on the intensity level instead of the amplitude level and system anisotropy manifests only in terms of multiplicative prefactors with the same scaling.\cite{Silbey2010,Chuang2014}

\begin{figure}[t]
	\centering
  \includegraphics[width=15.0cm]{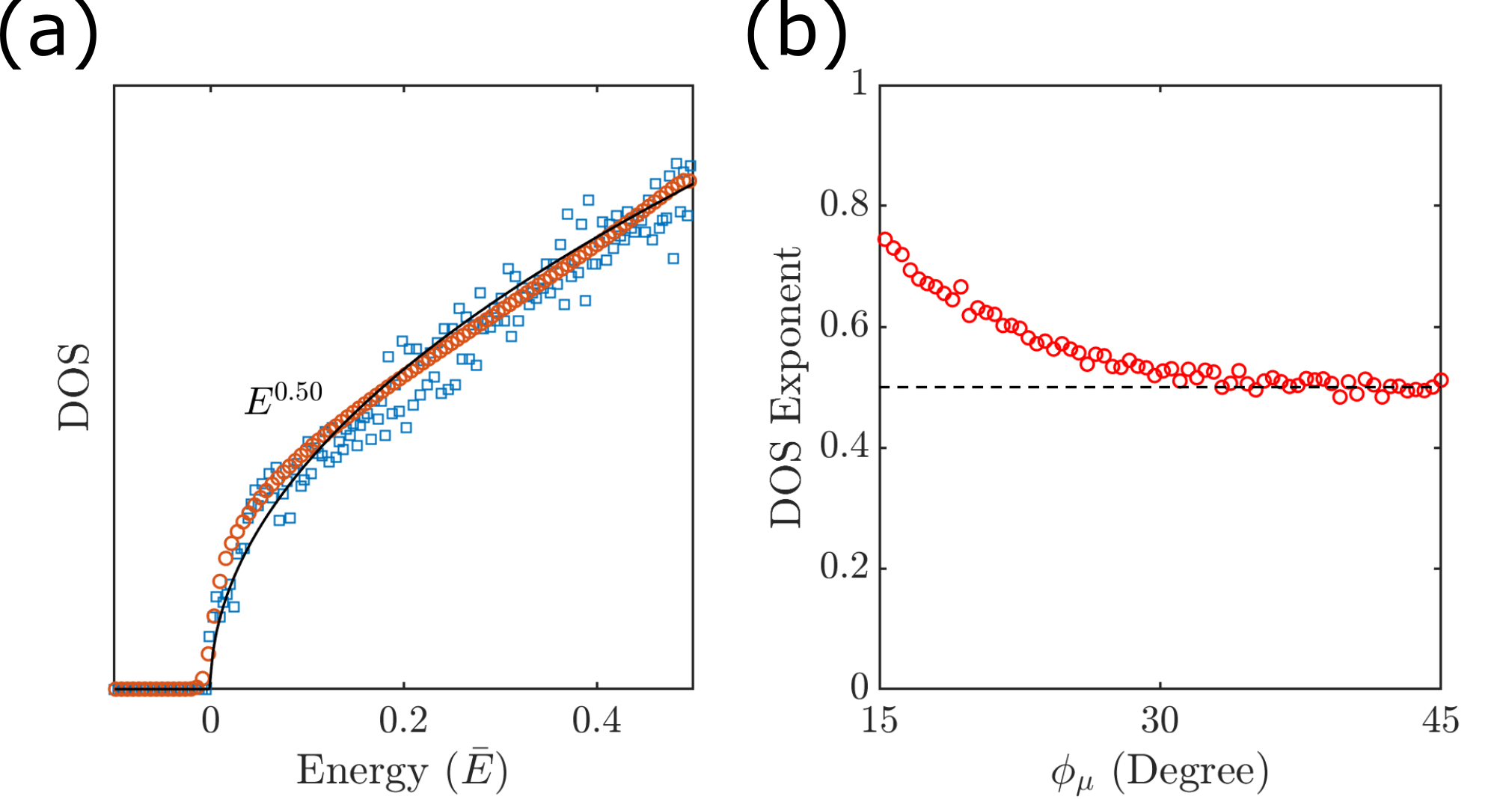}
  \caption{(a) The DOS from numerically evaluating Eq.~(\ref{eqn:continuum_exact}) (circles), that of a $\phi_\mu=\pi/4$ square lattice (squares), and a $E^{0.50}$ power-law fit (line). (b) Power-law exponents fitted for the low-energy DOS of square lattices with varying $\phi_\mu$. Only cases with the bright ($k=0$) state close to band edge are included.\cite{fn6} The dashed line indicates the theoretical value 0.5. }
  \label{fig:DOS}
\end{figure}

\textit{Universal scaling of density of states.}--- 
In order to connect from the universally scaled exciton dispersion to many of the spectroscopic features discussed below we need to ascertain the DOS near the bright state, particularly its scaling information when the bright state is at the band edge, \textit{i.e.} min$[E(\vec{k})]=E(0)$.\cite{fn4} We note that the DOS of an isotropically quadratic dispersion, \textit{e.g.} that of a 2D NN-coupled lattice, has a constant scaling ($E^{0}$). Also, the DOS of a 2D isotropically linear dispersion scales linearly ($E^{1}$). Consequently, the anisotropic dispersion predicted by Eq.~(\ref{eqn:continuum_exact}) leads to a DOS whose scaling is bounded by these two extrema. In fact we find that Eq.~(\ref{eqn:continuum_exact}) leads a $E^{\sim0.5}$ scaling of the DOS, as shown in Fig.~\ref{fig:DOS}(a) where the power-law fit agrees quantitatively up to $E-E(k=0)\approx0.5\bar{E}$.  An approximate derivation of $E^{\sim0.5}$ from the anisotropic dispersion is provided in S4 in SM. In Fig.~\ref{fig:DOS}(b) we show the fitted power-law exponent to the numerically calculated DOS of a dipole square lattice as a function of $\phi_\mu$. The deviation from $0.5$ in the $\phi_\mu<30^\circ$ regime is attributed to short-ranged discrete lattice contributions that dominate the large $k$ regime in the dispersion, whose energy overlaps with the small $k$ regime when $\phi_\mu$ is smaller. For $\phi_\mu>30^\circ$ the exponent converges to $0.5$.\cite{fn5} 

\textit{Spectroscopic signatures I: Disorder and temperature scalings of absorption linewidth.}---For typical 2D excitonic systems, observables in the optical regime ($k^{-1}\ge$100 nm) retrieving information are well described by the low $k$ dispersion Eq.~(\ref{eqn:continuum_k2}) and the corresponding $E^{0.5}$ scaling DOS. We explore a few such observables in the following sections.

\begin{figure}[t]
	\centering
  \includegraphics[width=11.0cm]{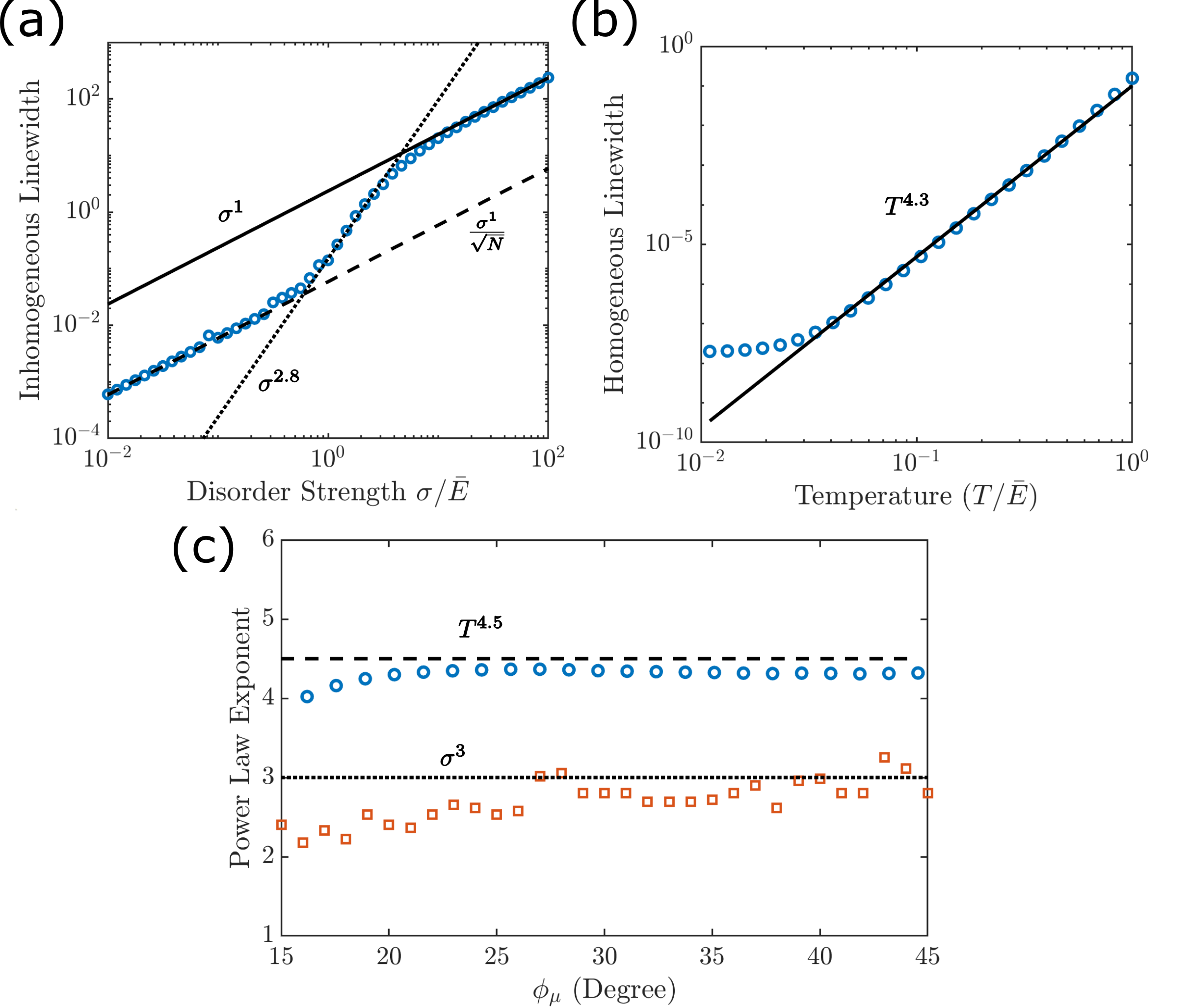}
  \caption{Disorder- and thermally induced absorption linewidths. (a) Absorption linewidth as a function of site disorder strength for a $\phi_\mu=\pi/4$ square lattice with 1600 sites. The solid and dashed lines indicate the strong and the weak disorder limits and the dotted line is a $\sigma^{2.8}$ fit to the intermediate regime. (b) Absorption linewidth as a function of temperature of the same system as in (a) coupled to a cubic super-Ohmic bath (dots) and its power-law fitting (line). (c) Power-law exponents of disorder (squares) and thermally (circles) induced absorption linewidths as functions of $\phi_\mu$ for square lattices. The horizontal lines indicate theoretically predicted $\sigma^3$ (dotted) and $T^{4.5}$ (dashed) scalings.}
  \label{fig:spectra}
\end{figure}

The influence of static disorder and thermal noise on the lineshape of excitonic systems in the condensed phase can be cast in terms of the DOS close to the bright states in the perturbative regime.\cite{SchreiberToyozawa1982,Jasper2005PRL,JianMa2015JCP,Chem2019} Specifically, it can be shown analytically that a $\mathrm{DOS}(E)\propto E^{0.5}$ scaling leads to a disorder-induced absorption width $W(\sigma)\propto\sigma^{3}$ using the coherent potential approximation (CPA).\cite{Huber1990JLumin,JasperPRL2015,AureliaCPA2017PRL} In Fig.\ref{fig:spectra}(a) we confirm the scaling by numerically computing the disorder-induced spectral width as a function of the strength of Gaussian uncorrelated site disorder.\cite{SchreiberToyozawa1982,Egorov1995CPL} In both the weak (fully delocalized, motionally narrowed\cite{Knapp1984}) and the strong disorder (fully localized) regimes the width scales linearly, while in the intermediate regime we find $W(\sigma)\propto\sigma^{2.8}$. By numerically solving the self-consistent equation of CPA with the dispersion Eq.~(\ref{eqn:continuum_k2}) we also obtain a scaling of $\sigma^{2.74}$ in agreement with both the analytic CPA ($\sigma^{3}$) and the exact ($\sigma^{2.8}$) results, detailed in S5 of SM.

In addition to the disorder-induced absorption width, the $E^{0.5}$ scaling of DOS also manifests in the temperature dependent spectral lineshape, a thermal noise-induced effect.\cite{MayKuhn} Utilizing the standard lineshape theory, Heijs et al. \cite{Jasper2005PRL} quantitatively explained the power-law temperature dependence of absorption width of linear PIC dye aggregates.\cite{Renge1997,Yamaguchi2006} Recently, basing on the same level of theory, we extended the analysis to temperature-dependent peak shift as a new means to characterize excitonic molecular solids.\cite{Chem2019} Under the fast bath assumption, \textit{i.e.} the bath relaxation time is faster than the inverse temperature, one arrives at power-law $T$-dependence for the linewidth:
\begin{eqnarray}
W(T)&=&\int_{E'}\mathrm{DOS}(E')\cdot J^\mathrm{(B)}(|E'-E(0)|)\cdot\bar{n}(E'-E(0),T)\nonumber\\
&=&W(0)+C_W\cdot T^{d_\mathrm{s}+s+1}
\end{eqnarray}
where $C_W$ is a constant and $\bar{n}(E,T)=(1-e^{-E/T})^{-1}$ is the Bose-Eistein distribution. $d_\mathrm{s}$ and $s$ are the power-law exponents of the system DOS and the bath spectral density $J^\mathrm{(B)}(\omega)$, respectively. For a cubic super-Ohmic bath, $J^\mathrm{(B)}(\omega)\propto\omega^s$ with $s=3$, we predict a $T^{4.5}$ dependence for 2D systems. In Fig.~\ref{fig:spectra}(b) we show the $T$-dependent linewidth of the $\phi_\mu=\pi/4$ square lattice. Both the disorder and the thermally induced absorption widths are in good agreements with their theoretically predicted counterparts, shown in Fig.~\ref{fig:spectra}(c) as well as in Fig. S7.\\

\textit{Spectroscopic signatures II: Absorption peak splittings.}---
The dispersion relation can be probed by scattering experiments such as the electron energy-loss spectroscopy that has become readily available for excitonic systems\cite{KnupferBerger2007PRL,KnupferBerger2012JCP}, and more recently using twisted light with designated orbital angular momenta exciton dispersions in the optical regime have also been experimentally measured.\cite{Lan2021ACSNano} On the other hand, a direct application of Eq.~(\ref{eqn:continuum_k2}) predicts the energy gap spanned by the asymmetric lineshape of a transient absorption spectrum.\cite{Mukamel1997} This gap measures the energy difference between the transition from the ground state to the one-particle bright state and that from the one-particle bright state to the two-particle bright state.

Taking the usual assumption that the two-particle bright state is dominated by the state $|\vec{k}_\mathrm{b}\rangle\otimes|2\vec{k}_\mathrm{b}\rangle$, the direct product of the one-particle bright state and the one-particle state of double quasimomentum,\cite{Dylan2013,Yoshida2020JPCL} with the bright state wavevector $\vec{k}_\mathrm{b}=k_b(-\sin\phi_\mu,\cos\phi_\mu)$, \textit{i.e.} perpendicular to the transition dipole moment. Ignoring two-exciton interactions, the splitting between the two peaks is:
\begin{eqnarray}
E(2\vec{k}_\mathrm{b})-E(\vec{k}_\mathrm{b})\approx-2+2\sin^2\theta_\mu\cdot(k_\mathrm{b}r_\mathrm{c})
\label{eqn:pumpprobegap}
\end{eqnarray} 
obtained by truncating Eq.~(\ref{eqn:continuum_k2}) to the linear order.

\begin{figure}[t]
	\centering
  \includegraphics[width=15cm]{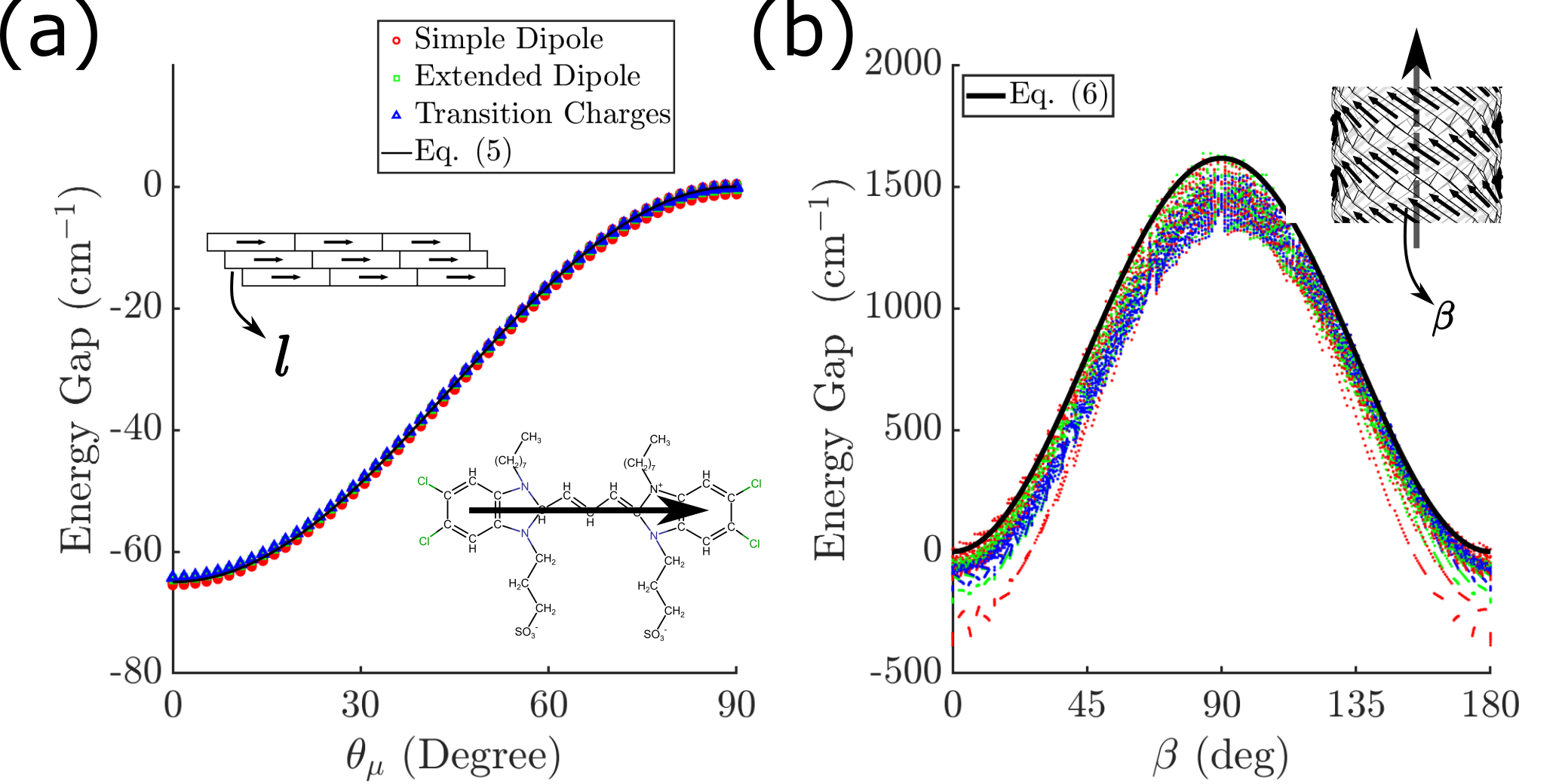}
  \caption{(a) The transient absorption gap of 2D molecular films and (b) the energy splitting between the perpendicular and the parallel-polarized absorption peaks of molecular tubules, compared to the theoretical predictions Eqs.~(\ref{eqn:pumpprobegap}) and (\ref{eqn:twopeaks}), respectively. The structure of C8S3 dye molecule is shown in  (a) and the arrow indicates the transition dipole direction of its lowest excited state. Here $r_\mathrm{c}=5$~\AA~and $k_\mathrm{b}=2\cdot10^{-4}$\AA$^{-1}$, corresponding to peak absorption at $500$ nm. Schematics of planar and tubular aggregates with their structural parameters (lattice offset $l$ for planes and helical pitch angle $\beta$ for tubes). We compare three different models of the excitonic coupling between two C8S3 molecules: simple dipole, extended dipole, and transition atomic charges. These models and the construction of planar and tubular lattices are detailed in S6 and S7 of SM.}
  \label{fig:gaps}
\end{figure}

Similar techniques can also be applied to tubular systems, where a common example are self-assembling amphiphilic dye molecules in solution.\cite{TubePRL,LucaTube2019NJP} In this case the system eigenstates can be labeled by $(k_\parallel,k_\perp)$ denoting the quasimomenta along the axial and the circumferential directions, respectively.\cite{LucaTube2019NJP} The selection rule dictates that bright states are those of $(k_\parallel,k_\perp)=(0,0)$ and $(0,\pm1)$. By analogy to the 2D planar system, the energy splitting between the two bright states to the linear order of $k_\perp\cdot r_\mathrm{c}$ can be written as
\begin{eqnarray}
E(0,\pm1)-E(0,0)=\frac{2\pi\mu_0^2}{ra_0}\sin^2\beta,
\label{eqn:twopeaks}
\end{eqnarray}
where $r$ is the tube radius and $\beta$ is the angle between the in-plane transition dipole and the tube axis, see inset of Fig.\ref{fig:gaps}. The detailed derivation and discussion of Eq.~(\ref{eqn:twopeaks}) is provided in S8 and S9 of SM. As predicted by Eq.~(\ref{eqn:twopeaks}), the gap is positive for all configurations with in-plane transition dipoles, \textit{i.e.}, the perpendicularly polarized peak is always higher in energy than the parallel one. We note that similar results for the energy gap and scaling exponents have been reported from numerical simulations of tubular systems.\cite{JasperPRL2015,Kohler2016}  The agreement between the 2D and tubular results is achieved only in the large radius limit, as tubular systems exhibit a 1D to 2D transition with increasing radius (see S9 and S11).\cite{TubePRL}  Thus, building on the anisotropic dispersion relation, we provide a systematic explanation of the numerical results in both 2D and tubular structures.

\textit{Application to excitonic molecular films and tubules.}---To corroborate our theoretical results and analysis including Eqs.~(\ref{eqn:pumpprobegap}) and (\ref{eqn:twopeaks}), we compare with the numerically evaluated exciton dispersions and the spectroscopic observables of model molecular films and tubules consisting of C8S3 dyes.\cite{Anu2019,Chem2019} The superstructures of the model aggregates studied include planar brick wall and helically symmetric tubular lattices with varying structural parameters, see the insets of Fig.~\ref{fig:gaps} (a) and (b) and S7 of SM. We further model the coupling matrix element $J(\vec{r}_n)$ with increasing spatial resolution on the molecular transition dipole distribution: from simple dipole and extended dipole to transition charges, as detailed in S6 of SM. As these methods differ only in the short range, we predict that the scaling properties in the small $k$ regime discussed above hold true for numerical results calculated from any of them. The quantitative agreements observed in Fig.~\ref{fig:gaps}(a) and (b) put Eq.~(\ref{eqn:continuum_k2}) to a good test. Additional examination of the power-law exponents of the DOS and $T$-dependent linewidth of these systems also show excellent agreement with those predicted by the continuum model, as discussed in S10 of SM, further substantiating the universality and applicability of the present theory.

\textit{Conclusion.}---By taking the continuum limit of 2D excitonic systems, we obtained an analytical expression of the anisotropic dispersions in the small $k$ regime where the continuum description is justified. We conclude that in this limit the exciton band scales linearly in the direction of the transition dipole and is dispersionless to linear order in the perpendicular direction, a result that can be understood by the quantum interference effects of interacting dipoles in 1D subsystems. Combining these features leads to the prediction of the $E^{\sim0.5}$ scaling of the 2D DOS near the bottom of the band and the explanation of a power-law disorder strength ($W(\sigma)\propto\sigma^{2.8}$) and temperature ($W(T)\propto T^{s+1.5}$) dependence of absorption linewidths, where $s$ is the characteristic scaling exponent of the bath degrees of freedom
. Expressions with $\sin^2\theta_\mu$ dependence for the energy splitting observable in transient absorption experiment for planar 2D systems and the energy gap between the two bright states of tubular systems are derived based on the anisotropic dispersion relation and can be directly applied to analyzing a large class of molecular systems. Our results are universal for 2D dipolar systems and provide a firm theoretical ground for understanding the photophysics of low-dimensional excitonic systems. \\

\begin{acknowledgments}
This work is supported by the NSF (CHE 1800301 and CHE 1836913).
\end{acknowledgments}

\begin{thebibliography}{57}
\bibitem [{fn1()}]{fn1}%
  \BibitemOpen
  \href@noop {} {\ }\bibinfo {note} {We note that a similar result was obtained
  in Ref.\cite{PhysRevB.57.11303}. However the authors stopped short at the
  linear order and did not discuss its implications (Eq.~(18) and (A19)
  therein).}\BibitemShut {Stop}%
\bibitem [{fn2()}]{fn2}%
  \BibitemOpen
  \href@noop {} {\ }\bibinfo {note} {While it is not always positive,
  Eq.~(\ref{eqn:continuum_k2}) is strictly concave up, \textit{i.e.} the energy
  at the origin, $k=0$, is the global minimum for $|k|\cdot
  r_\mathrm{c}\le16/5$.}\BibitemShut {Stop}%
\bibitem [{fn3()}]{fn3}%
  \BibitemOpen
  \href@noop {} {\ }\bibinfo {note} {The only parameters involved in
  Eqs.~(\ref{eqn:continuum_exact}) and (\ref{eqn:continuum_k2}) are the
  direction of the dipole relative to $\vec{k}$ ($\theta_\mu,\phi_\mu-\phi_k$),
  the area density ($\mu_0^2/a_0$), and the cutoff radius
  ($r_\mathrm{c}$).}\BibitemShut {Stop}%
\bibitem [{fn4()}]{fn4}%
  \BibitemOpen
  \href@noop {} {\ }\bibinfo {note} {In the regime where
  Eq.~(\ref{eqn:continuum_k2}) holds, there are two instances where the bright
  state $k=0$ is at the band edges: (1) $\theta_\mu=\pi/2$ such that
  min$[E(\vec{k})]=E(0)$ and (2) $\theta_\mu<\arccos(1/\sqrt{3})$ such that
  max$[E(\vec{k})]=E(0)$.\cite{Chem2019}}\BibitemShut {NoStop}%
\end{thebibliography}
%


\end{document}